\documentclass[10pt,twoside]{article}
\usepackage{amsmath,amssymb}

\makeatletter
\evensidemargin \oddsidemargin
\makeatother

\begin{document}
\thispagestyle{empty}
\setcounter{page}{1}



\begin{center}
{\large\bf SIMPLEX TRIANGULATION INDUCED SCALE-FREE NETWORKS}

\vskip.20in

Zhi-Ming Gu,$^{1}$\ Tao Zhou,$^{2,3,*}$\ Bing-Hong Wang,$^{2}$\ Gang Yan,$^{3}$\ Chen-Ping Zhu$^{1}$ \ and \ Zhong-Qian Fu$^{3}$\\[2mm]
{\footnotesize
$^{1}$College of Science, Nanjing University of Aeronautics and Astronautics\\
Nanjing Jiangsu, 210016, P. R. China\\[5pt]
$^{2}$Nonlinear Science Center and Department of Modern Physics\\
University of Science and Technology of China, Hefei Anhui, 230026, P. R. China\\[5pt]
$^{3}$Department of Electronic Science and Technology,\\
University of Science and Technology of China, Hefei Anhui, 230026, P. R. China\\[5pt]
}
\end{center}

{\footnotesize \noindent {\bf Abstract.}  We propose a simple rule
that generates scale-free networks with very large clustering
coefficient and very small average distance. These networks are
called simplex triangulation networks(STNs) as they can be
considered as a kind of network representation of simplex
triangulation. We obtain the analytic results of power-law
exponent $\gamma =2+\frac{1}{d-1}$ for $d$-dimensional STNs, and
clustering coefficient $C$. We prove that the increasing tendency
of average distance of STNs is a little slower than the logarithm
of the number of nodes in STNs. In addition, the STNs possess
hierarchical structure as $C(k)\sim k^{-1}$ when $k\gg d$ that in
accord with the observations of many real-life networks. \\[3pt]
{\bf Keywords.} complex networks, simplex triangulation, scale-free networks, small-world networks, clustering coefficient, average distance\\[3pt]
{\small\bf AMS (MOS) subject classification:} 05C75, 05C80}

\vskip.2in

\section{Introduction}

\noindent Recently, empirical studies indicate that the networks
in various fields have some common characteristics, which inspires
scientists to construct a general model[1-3]. The most important
characteristics are scale-free property and small-world effect.
The former means that the degree distribution obeys power law as
$p(k)\propto k^{-\gamma }$, where $k$ is the degree and $p(k)$ is
the probability density function for the degree distribution.
$\gamma$ is called the power-law exponent, and is usually between
2 and 3 in real-life networks. The latter involves two factors:
small average distance as $L\sim \texttt{ln}N$ or even smaller and
great clustering coefficient $C$, where $L$ is the average
distance, $N$ is the number of nodes in the network, and $C$ is
the probability that a randomly selected node's two randomly
picked neighbors are neighbor. One of the most well-known models
is Watts and Strogatz's small-world network (WS network), which
can be constructed by starting with a regular network and randomly
moving one endpoint of each edge with probability $p$[4]. Another
significant one is Barab\'{a}si and Albert's scale-free network
model (BA network)[5]. The BA model suggests that two main
ingredients of self-organization of a network in a scale-free
structure are growth and preferential attachment. However, both WS
and BA networks fail to mimic the reality in some aspects.
Therefore, a significant problem is how to generate networks
displaying both scale-free property and small-world effect[6-8].

In this paper, we propose a simple rule that generates scale-free
networks with very large clustering coefficient and very small
average distance. These networks are called simplex triangulation
networks(STNs) as they can be considered as the one-dimensional
framework of simplex triangulation in the view of algebraic
topology[9]. Strictly speaking, if
$\vec{a_0},\vec{a_1},\cdots,\vec{a_d}$ are the linear independent
points in $\mathbb{R}^d$, the set
$\{\sum^d_{i=0}\lambda_i\vec{a_i} | \lambda_i\geq 0,
\sum^d_{i=0}\lambda_i=1 \}$ is the $d$-simplex[9]. For instance,
0-simplex is a single vertex, 1-simplex is a line segment,
2-simplex is a triangle, 3-simplex is a tetrahedron, and so on.
Informal speaking, simplex triangulation is the process to
triangulate original simplex into several sub-simplices. For
example, choose an arbitrary point inside a $d$-simplex, , and
link this point to all the vertices of this simplex, then this
simplex will be triangulated into $d+1$ sub-simplices.

Our model starts with a $d$-simplex, where $d\geq 2$ is an
integer. The network representation of this simplex contains $d+1$
nodes and $\frac{d(d+1)}{2}$ edges. Then, at each time step, a
simplex is randomly selected, and a new node is added inside this
simplex and linked to all its $d+1$ vertices(nodes). During each
time step, the number of simplices, nodes and edges increases by
$d$, one, $d+1$, respectively. Repeat this simple rule, one can
get the $d$-STN of arbitrary order that he likes.

\section{The degree distribution}

\noindent As we have mentioned above, the degree distribution is
one of the most important statistical characteristics of networks.
Since many real-life networks are scale-free networks, whether the
networks are of power-law degree distribution is a criterion to
judge the validity of the model.

Since after a new node is added to the network, the number of
simplices increases by $d$, we can immediately get that when the
network is of order $N$, the number of simplices is:
\begin{equation}
N_s=d(N-d-1)+1=dN-d^2-d+1
\end{equation}
Let $N_s^i$ be the number of simplices containing the node $i$,
the probability that a newly added node will link to the $i$th
node is $N_s^i/N_s$. Apparently, $N_s^i$ will increase by $d-1$
while $k_i$ increases by one, and when the node is newly added, it
is of degree $d+1$ and contained by $d+1$ different simplices,
thus
\begin{equation}
N_s(k)=k(d-1)-(d+1)(d-2),
\end{equation}
where $N_s(k)$ denotes the number of simplices containing a node
of degree $k$. Let $n(N,k)$ be the number of nodes with degree $k$
when $N$ nodes are present, now we add a new node to the network,
$n(N,k)$ evolves according to the following rate equation[10]:
\begin{equation}
n(N+1,k+1)=n(N,k)\frac{N_s(k)}{N_s}+n(N,k+1)(1-\frac{N_s(k+1)}{N_s})
\end{equation}
When $N$ is sufficient large, $n(N,k)$ can be approximated as
$Np(k)$, where $p(k)$ is the probability density function for the
degree distribution. In terms of $p(k)$, the above equation can be
rewritten as:
\begin{equation}
p(k+1)=\frac{N}{N_s}[p(k)N_s(k)-p(k+1)N_s(k+1)]
\end{equation}
Using Eq.(1) \& (2) and the expression
$p(k+1)-p(k)=\frac{dp}{dk}$, we can get the continuous form of
Eq.(4):
\begin{equation}
(2d-1-\frac{d^2+d-1}{N})p(k+1)+[k(d-1)-(d+1)(d-2)]\frac{dp}{dk}=0
\end{equation}
For large $k$ ($p(k+1)\approx p(k)$, $k\gg d$) and $N\gg d$, we
have:
\begin{equation}
(2d-1)p(k)+(d-1)k\frac{dp}{dk}=0
\end{equation}
This lead to $p(k)\propto k^{-\gamma}$ with
$\gamma=2+\frac{1}{d-1}\in (2,3]$, in accord with many real-life
networks with power exponent between 2 and 3.

\section{The clustering coefficient}
\noindent The clustering coefficient of a node is the ratio of the
total number of existing edges between all its neighbors and the
number of all possible edges between them. The clustering
coefficient $C$ of the whole network is defined as the average of
all nodes' clustering coefficient. At first, let us derive the
analytical expression of $C(k)$ denoting the clustering
coefficient of a node with degree $k$. When a node is newly added
into the network, its degree and clustering coefficient are $d+1$
and 1. After that, if its degree increases by one, then its new
neighbor must link to its $d$ existing neighbors. Hence we have:
\begin{equation}
C(k)=\frac{\frac{d(d+1)}{2}+d(k-d-1)}{\frac{k(k-1)}{2}}=\frac{d(2k-d-1)}{k(k-1)}
\end{equation}

The clustering coefficient of the whole network can be obtained as
the mean value of $C(k)$ with respect to the degree distribution
$p(k)$:
\begin{equation}
C=\int_{k_{\texttt{min}}}^{k_{\texttt{max}}}C(k)p(k)dk,
\end{equation}
where $k_{\texttt{min}}=d+1$ is the minimal degree and
$k_{\texttt{max}}\gg k_{\texttt{min}}$ is the maximal degree.
Combine Eq.(7) and Eq.(8), note that $p(k)=Ak^{\frac{2d-1}{d-1}}$
and $\int_{k_{\texttt{min}}}^{k_{\texttt{max}}}Ap(k)dk=1$, we can
get the analytical result of $C$ by approximately treating
$k_{\texttt{max}}$ as $+\infty$. For example, when $d=2$, the
clustering coefficient is
\begin{equation}
C=\frac{46}{3}-36\texttt{ln}\frac{3}{2}\approx 0.7366,
\end{equation}
and when $d=3$, it is
\begin{equation}
C=18+36\sqrt{2}\texttt{arctan} \sqrt{2}+\frac{9}{2}\pi
-18\sqrt{2}\pi \approx 0.8021
\end{equation}
For larger $d$, the expression is too long thus it will not be
shown here. The integral values for $d=4,5,6,7,8,9,10$ are 0.8406,
0.8660, 0.8842, 0.8978, 0.9085, 0.9171, and 0.9241, respectively.

It is remarkable that, the clustering coefficient of BA networks
is very small and decreases with the increasing of network order,
following approximately $C\sim \frac{\texttt{ln}^2N}{N}$[11].
Since the data-flow patterns show a large amount of clustering in
interconnection networks, the STNs may perform better than BA
networks. In addition, the demonstration exhibits that most
real-life networks have large clustering coefficient no matter how
many nodes they have. That agrees with the case of STNs but
conflicts with that of BA networks. Further more, many real-life
networks are characterized by the existence of hierarchical
structure[12], which can usually be detected by the negative
correlation between the clustering coefficient and the degree. The
BA network, which does not possess hierarchical structure, is
known to have the clustering coefficient $C(x)$ of node $x$
independent of its degree $k(x)$, while the STN has been shown to
have $C(k)\sim k^{-1}$, in accord with the observations of many
real networks[12].

\section{The average distance}
\noindent Using symbol $d(i,j)$ to represent the distance between
$i$ and $j$, the average distance of STN with order $N$, denoted
by $L(N)$, is defined as: $L(N)=\frac{2\sigma (N)}{N(N-1)}$, where
the total distance is: $\sigma (N)=\sum_{1\leq i<j\leq N}d(i,j)$.
Since newly added node will not affect the distance between
existing nodes, we have:
\begin{equation}
\sigma(N+1)=\sigma(N)+\sum_{i=1}^N d(i,N+1)
\end{equation}
Assume that the node $N+1$ is added into the simplex $y_1y_2\cdots
y_{d+1}$, then the Eq.(11) can be rewritten as:
\begin{equation}
\sigma(N+1)=\sigma(N)+\sum_{i=1}^N (D(i,y)+1)
\end{equation}
where $D(i,y)=\texttt{min}\{d(i,y_j),j=1,2,\cdots,d+1\}$.
Constrict this simplex continuously into a single node $y$, then
we have $D(i,y)=d(i,y)$. Since $d(y_j,y)=0$, Eq.(12) can be
rewritten as:
\begin{equation}
\sigma(N+1)=\sigma(N)+N+\sum_{i\in \Gamma} d(i,y)
\end{equation}
where $\Gamma =\{1,2,\cdots ,N\}-\{y_1,y_2,\cdots ,y_{d+1}\}$ is a
node set with cardinality $N-d-1$. The sum $\sum_{i\in \Gamma}
d(i,y)$ can be considered as the total distance from one node $y$
to all the other nodes in STN with order $N-d$. In a rough
version, the sum $\sum_{i\in \Gamma} d(i,y)$ is approximated in
terms of $L(N-d)$ as:
\begin{equation}
\sum_{i\in \Gamma} d(i,y)\approx (N-d-1)L(N-d)
\end{equation}
Apparently,
\begin{equation}
(N-d-1)L(N-d)=\frac{2\sigma(N-d)}{N-d}<\frac{2\sigma(N)}{N}
\end{equation}
Combining Eqs.(13), (14) and (15), one can obtain the inequality:
\begin{equation}
\sigma(N+1)<\sigma(N)+N+\frac{2\sigma(N)}{N}
\end{equation}
Consider (15) as an equation, then we have:
\begin{equation}
\frac{d\sigma(N)}{dN}=N+\frac{2\sigma(N)}{N}
\end{equation}
This equation leads to
\begin{equation}
\sigma(N)=N^2\texttt{ln}N+H
\end{equation}
where $H$ is a constant. As $\sigma(N)\sim N^2L(N)$, we have
$L(N)\sim \texttt{ln}N$. Which should be pay attention to, since
Eq. (16) is an inequality indeed, the precise increasing tendency
of $L$ may be a little slower than $\texttt{ln}N$.

\section{Conclusion remarks}
\noindent In conclusion, in respect that the simplex triangulation
networks are of very large clustering coefficient and very small
average distance, they are not only the scale-free networks, but
also small-world networks. Since many real-life networks are both
scale-free and small-world, STNs may perform better in mimicking
reality than BA and WS networks. In addition, STNs possess
hierarchical structure in accord with the observations of many
real networks.

Further more, we propose an analytic approach to calculate
clustering coefficient, and an interesting technique to estimate
the growing trend of average distance. Since in the earlier
studies, only few analytic results about these two quantities of
networks with randomicity are reported, we believe that our work
may enlightened readers on this subject.

We also have done the corresponding numerical simulations, the
simulation results agree with the analytical ones very well.

\section{Acknowledgements}

\noindent ZMGu and CPZhu acknowledges the support by the National
Natural Science Foundation of China(NNSFC)under No. XXXXXXXX.
BHWang acknowledges the support by NNSFC under No. 70271070 and
No. 10472116, and by the Specialized Research Fund for the
Doctoral Program of Higher Education under No. 20020358009. TZhou
acknowledges the support by NNSFC under No. 70471033, and the
Foundation for Graduate Students of University of Science and
Technology of China under Grant No. KD200408.

\footnotesize
\section{References}

\begin{itemize}

\item[{[1]}]
R. Albert and A. -L. Barab\'{a}si, Statistical mechanics of
complex networks, {\em Rev. Mod. Phys.}, {\bf 74}, (2002) 47-97.

\item[{[2]}]
M. E. J. Newman, The structure and function of
complex networks, {\em SIAM Rev.}, {\bf 45}, (2003) 167-256.

\item[{[3]}]
X. -F. Wang, Complex networks: topology, dynamics and
synchronization, {\em Int. J. Bifurcation \& Chaos}, {\bf 12},
(2002) 885-916.

\item[{[4]}]
D. J. Watts and S. H. Strogatz, Collective dynamics
of small-world networks, {\em Nature}, {\bf 393}, (1998) 440-442.

\item[{[5]}]
A. -L. Barab\'{a}si and R. Albert, Emergence of
scaling in random networks, {\em Science}, {\bf 286}, (1999)
509-512.

\item[{[6]}]
P. Holme and B. J. Kim, Growing scale-free networks
with tunable clustering, {\em Phys. Rev. E}, {\bf 65}, (2002)
026107.

\item[{[7]}]
J. S. Andrade Jr., H. J. Herrmann, R. F. S. Andrade
and L. R. da Silva, Apollonian Networks: Simultaneously
scale-free, small world, Euclidean, space filling, and with
matching graphs. {\em Phys. Rev. Lett.}, {\bf 94}, (2005) 018702.

\item[{[8]}]
Tao Zhou, Gang Yan and Bing-Hong Wang, Maximal planar
networks with large clustering coefficient and power-law degree
distribution, {\em Phys. Rev. E}, {\bf 71}, (2005) 046141.

\item[{[9]}] J. R. Munkres, {\em Elements of Algebraic Topology},
Addison-Wesley Publishing Company, 1984.

\item[{[10]}]
P. L. Krapivsky, S. Redner and F. Leyvraz,
Connectivity of growing random networks, {\em Phys. Rev. Lett.},
{\bf 85}, (2000) 4629-4632.

\item[{[11]}] K. Klemm and V. M. Egu\'{i}luz, High clustered
scale-free networks, {\em Phys. Rev. E}, {\bf 65}, (2002) 036123.

\item[{[12]}] E. Ravasz and A. -L. Barab\'{a}si, Hierarchical
organization in complex networks, {\em Phys. Rev. E}, {\bf 67},
(2003) 026112.

\end{itemize}

\vfill

\noindent $^{*}$ corresponding author, email:\ zhutou@ustc.edu

\end{document}